\documentclass[pdflatex,sn-mathphys-num]{sn-jnl}

\usepackage{graphicx}%
\usepackage{multirow}%
\usepackage{amsmath,amssymb,amsfonts}%
\usepackage{amsthm}%
\usepackage{mathrsfs}%
\usepackage[title]{appendix}%
\usepackage{xcolor}%
\usepackage{textcomp}%
\usepackage{manyfoot}%
\usepackage{booktabs}%
\usepackage{algorithm}%
\usepackage{algorithmicx}%
\usepackage{algpseudocode}%
\usepackage{listings}%
\usepackage{subfig}
\usepackage{comment}
\usepackage{dsfont}
\usepackage{hyperref}

\theoremstyle{thmstyleone}%
\theoremstyle{thmstyletwo}%

\theoremstyle{thmstylethree}%

\begin{document}

\title[ProstNFound+]{ProstNFound+: A Prospective Study using Medical Foundation Models for Prostate Cancer Detection}


\author*[1]{\fnm{Paul F.R.} \sur{Wilson}}\email{paul.wilson@queensu.ca}

\author[1]{\fnm{Mohamed} \sur{Harmanani}}
\author[2]{\fnm{Minh Nguyen Nhat} \sur{To}}
\author[1]{\fnm{Amoon} \sur{Jamzad}}
\author[2]{\fnm{Tarek} \sur{Elghareb}}
\author[1]{\fnm{Zhuoxin} \sur{Guo}}
\author[3]{\fnm{Adam} \sur{Kinnaird}}
\author[4]{\fnm{Brian} \sur{Wodlinger}}

\author[2]{\fnm{Purang} \sur{Abolmaesumi}}
\equalcont{These authors contributed equally to this work.}

\author[1]{\fnm{Parvin} \sur{Mousavi}}
\equalcont{These authors contributed equally to this work.}

\affil[1]{\orgname{Queen's University}, \city{Kingston},  \country{Canada}}

\affil[2]{\orgname{University of British Columbia}, \city{Vancouver},  \country{Canada}}

\affil[3]{\orgname{University of Alberta}, \city{Edmonton},  \country{Canada}}

\affil[4]{\orgname{Exact Imaging}, \city{Markham},  \country{Canada}}



\abstract{The abstract serves both as a general introduction to the topic and as a brief, non-technical summary of the main results and their implications. Authors are advised to check the author instructions for the journal they are submitting to for word limits and if structural elements like subheadings, citations, or equations are permitted.}


\abstract{

\textbf{Purpose:} Medical foundation models (FMs) offer a path to build high-performance diagnostic systems. However, their application to prostate cancer (PCa) detection from micro-ultrasound ($\mu$US) remains untested in clinical settings. We present ProstNFound+, an adaptation of FMs for PCa detection from $\mu$US, along with its first prospective validation. 
\textbf{Methods:}
ProstNFound+ incorporates a medical FM, adapter tuning, and a custom prompt encoder that embeds PCa-specific clinical biomarkers. The model generates a cancer heatmap and a risk score for clinically significant PCa. Following training on multi-center retrospective data, the model is prospectively evaluated on data acquired five years later from a new clinical site. Model predictions are benchmarked against standard clinical scoring protocols (PRI-MUS and PI-RADS). \textbf{Results:}
ProstNFound+ shows strong generalization to the prospective data, with no performance degradation compared to retrospective evaluation. It aligns closely with clinical scores and produces interpretable heatmaps consistent with biopsy-confirmed lesions. 
\textbf{Conclusion:}
The results highlight its potential for clinical deployment, offering a scalable and interpretable alternative to expert-driven protocols.\footnote{Training and evaluation code is publicly available at \href{https://github.com/pfrwilson/prostNfound}{github.com/pfrwilson/prostNfound}}}



\maketitle

\section{Introduction}\label{sec1}

Prostate cancer (PCa) is a leading cause of cancer-related death among men worldwide, with its burden expected to rise due to an aging global population.
Early detection of PCa, especially \emph{clinically significant} cancers (csPCa) which are aggressive and require immediate treatment, is critical for achieving favorable outcomes. Imaging plays a key role in PCa diagnosis, with visual risk assessment systems enabling the identification of suspicious lesions from images and the assessment of their risk. Identifying lesions facilitates accurate biopsy for histopathological analysis and conclusive diagnosis~\cite{ahmed2017diagnostic}. Meanwhile, risk scoring to differentiate between csPCa and clinically insignificant (isPCa) can allow patients with low-risk lesions to avoid biopsy and potential overdiagnosis, while ensuring high-risk lesions are targeted during biopsy~\cite{turkbey2023pi}.

The most-established visual risk assessment system is PI-RADS, based on the analysis of pre-procedural multiparametric magnetric resonance imaging (mpMRI) of the prostate~\cite{ahmed2017diagnostic}. The high cost of mpMRI and the need for dedicated facilities limit its widespread global adoption; hence, more cost-effective, convenient and scalable imaging methods are needed. High-resolution micro-US ($\mu$US) provides a promising alternative, with the PRI-MUS scoring system providing risk assessment based on characteristic echotextural features associated with the prostate tissue. A recent randomized clinical trial with 678 participants reported that $\mu$US-targeted biopsy is non-inferior to mpMRI-targeted biopsy for csPCa diagnosis~\cite{kinnaird2025microultrasonography}. Despite its success, the PRI-MUS protocol depends on operator expertise, requires extensive training, and can suffer from significant inter-observer variability. Consequently, there is a need for a complementary, user-agnostic, and objective cancer detection tool using $\mu$US to standardize lesion targeting and reliably detect csPCa across diverse clinical settings.

Deep learning methods have shown promising success in various medical imaging tasks including 
PCa detection using mpMRI~\cite{fassia2024deep}, conventional US~\cite{sun2023three}, temporal-enhanced US~\cite{fooladgar2022uncertainty}, and $\mu$US~\cite{pensa2024deep}. The dominant paradigm has involved developing relatively small architectures, typically based on convolutional network networks, trained from scratch on single-modality datasets. The limited size of available $\mu$US datasets and the tendency of neural networks to exhibit poor generalization when trained on small data have presented major challenges.
Medical foundation models (FMs), i.e., large neural networks pretrained on diverse datasets, have emerged as a transformative paradigm for medical imaging. 
 FMs have demonstrated state-of-the-art (SOTA) performance across diverse tasks in radiology~\cite{alzate2023sam}, 
 including those for interpretation of conventional US images~\cite{jiao2024usfm}. 
FMs learn rich feature representations during pretraining that are transferred to downstream applications, reducing the need for large annotated datasets. In practice, several domain gaps limit their application for specialized clinical tasks such as PCa detection.  FMs that have not been trained on $\mu$US lack understanding its specific features. Additionally,  models developed for standard segmentation tasks are not immediately well suited for PCa assessment, which requires identifying subtle differences in tissue appearance without clear boundaries. Finally,  existing FMs do not explicitly incorporate clinical context relevant to PCa. 

To address domain gaps, adaptation methods have been explored in the literature. Finetuning strategies such as full finetuning
~\cite{ma2024segment} 
and prompt tuning
~\cite{fischer2024prompt} have been used to adapt natural imaging FMs to the medical domain. Architectural adaptations have also been proposed, including the addition of convolutional branches to better capture texture information in conventional US~\cite{lin2023samus} or high-resolution features in MRI~\cite{alzate2023sam}. Conditional prompting~\cite{zhou2022conditional} using auxiliary imaging or non-imaging data to generate prompts that guide FM output have been explored for FM adaptation~\cite{gaillochet2024automating}.  
Recognizing the need for strong and flexible adaptation methods, we proposed ProstNFound, a unified model that combined finetuning with architectural modifications and conditional prompting with PCa specific clinical biomarkers to achieve SOTA performance in PCa detection from $\mu$US. 

Despite these advances, the real-world value of deep learning systems for PCa detection remains largely unverified for several reasons: 
(i) most existing work focuses on PCa detection not csPCa risk assessment or differentiation between csPCa and isPCa, critical for clinical decision making;  (ii) previous studies do not directly benchmark against established visual protocols, limiting a clear insight on true clinical impact; and (iii) mirroring a broader trend in medical deep learning~\cite{liu2019comparison}, methods are evaluated on retrospective data, risking overoptimistic performance estimates. To assess clinical performance, where shifts in patient populations and imaging protocols over time can cause unexpected degradation, prospective evaluation is key.

Extending our previous study ProstNFound~\cite{wilson2024prostnfound}, we introduce ProstNFound+, an accurate and generalizable AI model for PCa detection and csPCa risk scoring using $\mu$US. ProstNFound+ adapts medical FMs through adapter layers, a customized prompt encoder network designed to embed clinical biomarker data, and a multi-head output module that simultaneously predicts a cancer heatmap and a csPCa risk score. We present rigorous prospective validation and comparison to visual scoring systems. Our key novel contributions are:

\begin{enumerate}
    \item expanding the original ProstNFound architecture by adding a multi-head output module that simultaneously predicts a cancer localization heatmap and a patient-level risk score for csPCa. This design facilitates both regional interpretation for targeted biopsy and global risk assessment for clinical decision-making. ProstNFound+ outperforms prior SOTA in deep learning PCa detection for $\mu$US.
    \item presenting, for the first time, a prospective evaluation of deep learning models for $\mu$US-based PCa detection. Our model successfully generalizes to a prospective dataset collected from a new clinical center five years after the training dataset.
    \item  performing the first study (to the best of our knowledge) that benchmarks a deep learning model against PRI-MUS risk scores. We report competitive performance, with a relatively small (-5\% AUROC) gap compared to highly-trained human experts, demonstrating our model's potential for clinical translation.
\end{enumerate}

\section{Materials}\label{sec1}


Our study involves two datasets: a \textit{\textbf{retrospective}} training and cross-validation dataset and a \textbf{\textit{prospective}} test dataset. Patients underwent transrectal US-guided prostate biopsy using the ExactVu $\mu$US system (Exact Imaging, Markham, Canada). The system employs a high-frequency (29~MHz), side-firing linear-array transrectal probe (512 elements, $90~\mu m$ pitch). 

\textbf{Images and Labels:} B-mode $\mu$US images in the sagittal plane, with a depth of $28~mm$ and width of $46.06~mm$, were recorded during biopsy procedures. For each biopsy sample, the region corresponding to the needle trace was annotated, and the frame immediately preceding needle firing was included in the dataset. Histopathology determined PCa diagnosis, cancer involvement (percentage of cancer within the tissue sample), and ISUP grade group (GG), a score ranging from 0 (benign) to 5 (highly aggressive cancer) of biopsies. Samples were labeled as csPCa if their GG $>= 3$, and non-csPCa otherwise. Non-csPCa was further subdivided into No cancer (GG=0) and isPCa (GG=1 or 2). The choice of GG $>=3$ as the definition for csPCa follows~\cite{ahmed2017diagnostic} and corresponds to PCa of at least intermediate risk. Subjects were assigned an overall label of No cancer, isPCa or csPCa based on their most clinically significant biopsy core. Pathology labels were assigned directly to the corresponding annotated needle trace regions on the image. Other clinical metadata includes prostate-specific antigen (PSA), age, approximate PSA density (PSAD) and family history.

\textbf{Retrospective Data} are from a clinical trial conducted between 2013 and 2016, involving five centers (\href{https://clinicaltrials.gov/study/NCT02079025}{NCT02079025}). Patients underwent systematic (untargeted) biopsy of 10-12 cores per patient. 
The trial predated the development of the PRI-MUS protocol, hence risk scores are not available. In total, there are 693 subjects and 6607 biopsy cores, of which 5727 (86\%) were benign, 480 (7\%) isPCa, and 400 (6\%) csPCa. 


\textbf{Prospective Data} are from the OPTIMUM clinical trial conducted between 2021 and 2024 (\href{https://clinicaltrials.gov/study/NCT05220501}{NCT05220501})~\cite{kinnaird2025microultrasonography}, specifically subjects from the clinical center located in Edmonton, Canada. Each patient underwent a systematic and targeted biopsy, the latter guided either by $\mu$US-based PRI-MUS scores or preprocedural mpMRI-based PI-RADS scores, depending on the trial arm. PRI-MUS or PI-RADS scores corresponding to each biopsy sample are provided. The dataset includes 77 subjects and 1040 biopsy cores, of which 672 (64\%) were benign, 162 (25\%) isPCa, and 105 (10\%) csPCa. \emph{We remained blinded to all data and pathology results} until the completion of model training, at which point the final ProstNFound+ model was fixed. 

\textbf{Preprocessing:} B-mode images were resized from $1372 \times 833$ pixels to $256 \times 256$ pixels using bilinear interpolation.\footnote{Improvements in training and inference speeds occurred at this lower resolution with no significant decrease in model performance.} Pixel values were scaled to the range $(0, 1)$. Clinical metadata age, PSA, and PSAD were normalized using statistics of retrospective data.

\section{Methods}

\begin{figure*}[t]
    \centering
    \includegraphics[width=.95\textwidth]{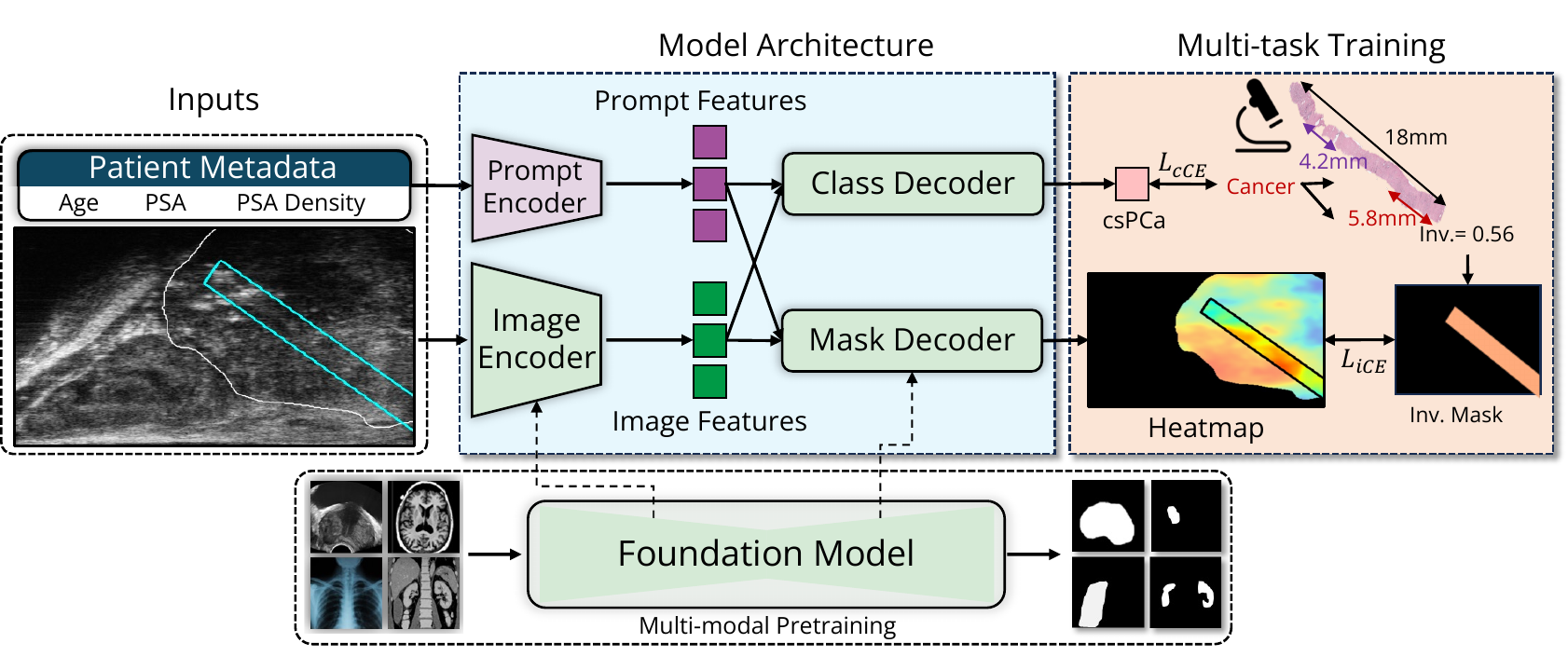}
    \caption{ProstNFound+ integrates a B-mode image encoder with conditional prompting using clinical metadata. The resulting embeddings are used by the mask decoder to generate a heatmap of cancer likelihood, and by the class decoder to output an image-level score representing the likelihood of clinically significant prostate cancer (csPCa).}
    \label{fig:pnf_mainfig}
\end{figure*}

An overview of our method is provided in Figure~\ref{fig:pnf_mainfig}. Our model, built by adapting a medical FM, takes B-mode $\mu$US images and patient metadata as inputs, producing a csPCa \emph{model risk score} and a cancer heatmap. 

\textbf{Foundation Model:} Our model borrows the image encoder and mask decoder components from the MedSAM~\cite{ma2024segment} medical segmentation foundation model. The encoder ($\approx$ 90M parameters) is a vision transformer and extracts $256\times64\times64$ embeddings of B-mode images. The mask decoder ($\approx 6M$ parameters) accepts the image embeddings along with zero or more 256-dimensional prompt embeddings to generate a $256\times256$-dimensional heatmap $\hat{Y}_\text{hmap}$. It consists of a sequence of attention blocks between image and prompt embeddings, followed by transposed convolutions.

\textbf{Conditional Prompting:}
Following the original ProstNFound~\cite{wilson2023self} method, we design an auxiliary prompt encoder network that allows us to condition the foundation model on PCa-specific clinical context. The network embeds clincal markers of age, PSA, and PSAD using two-layer MLPs to project the markers' scalar values to 256 dimensions and outputs $N$x256 ( $N$ is the number of prompts) embeddings.

\textbf{Multi-Head Architecture:} Visual scoring systems like PRI-MUS support not only the detection of PCa lesions, but also the identification of cancers that are likely to be clinically significant (high grade and aggressive). To support this important usage scenario, we add a secondary \emph{class decoder} head in addition to the existing mask decoder. The class decoder accepts the image and prompt embeddings and generates a scalar value $\hat{Y}_\text{csPCa}\in [0, 1]$ representing csPCa risk. It uses the same architecture as the mask decoder, but uses a linear layer rather than deconvolution layers.

\textbf{Multi-Task Loss Function:} We train the model in a two-term, multi-task fashion, simultaneously optimizing its heatmap and csPCa outputs and using pathology labels as ground truth. The first term, the \emph{csPCa}, encourages the model's class decoder output to match the true finding of csPCa or non-csPCa for a biopsy sample using cross-entropy: $\mathcal{L}_\text{csPCa} = H(Y_\text{csPCa}, \hat{Y}_\text{csPCa})$, 
where $Y_\text{csPCa}$ is the true csPCa label and $H(\cdot, \cdot)$ is the cross-entropy. The second term, \emph{heatmap}, encourages the model's output pixel values to be high for PCa regions and low for others. Since pixel-level PCa annotations are unavailable, we follow~\cite{harmanani2025cinepro} and use a loss function based on the involvement of cancer in the sample. For a given heatmap score $\hat{Y}_\text{hmap}$, the ``predicted involvement" is the average heatmap activation in the needle region, i.e., $\hat{I} = \frac{1}{|\mathcal{N}|} \sum_{i, j \in N} X_{i,j}$, where $\mathcal{N}$ denotes the set of pixel indices in the needle mask. Given the true involvement $I$, the loss is $H(I, \hat{I})$. The total loss is given by the sum $\mathcal{L} = \mathcal{L}_\text{csPCa} + \mathcal{L}_\text{hmap}$.

\textbf{Model Risk Score:} To allow comparison with PRI-MUS and PI-RADS (both a scale of 1–5), we discretize the outputs of the csPCa head into a \emph{model risk score} from 1 to 5. The outputs $\hat{Y}_\text{csPCa}\in [0, 1]$ are discretized by partitioning the interval $[0, 1]$ into bins $[0, t_1), [t_1, t_2), ..., [t_4, 1]$ and assigning $\hat{Y}_\text{csPCa}$ a score based on the bin it falls into. The bins (and thresholds $t_i$) are chosen so that the empirical distribution of \emph{model risk scores} follows that of PRI-MUS scores, using histogram matching.

\section{Experiments}

Our experimental setup is divided into two main phases: (1) training and validation on retrospective multi-center data, and (2) evaluation on a held-out prospective test set. 

\textbf{Training and Development:} In this phase we aim to develop and choose the best performing model on the retrospective data. We use a five-fold cross validation scheme, where the retrospective dataset is split by subjects into five non-overlapping folds. We perform each experiment five times, each time using a different fold as the validation set, and average the performance metrics, providing a robust estimate of the models' ability to generalize. \textit{Baselines:} We implement and test relevant baselines from the literature. First, we evaluate patch-based classification using ResNet~\cite{wilson2023self} and a finetuned MicroSegNet~\cite{jiang2024microsegnet}. For comparison with other foundation model strategies, we re-implemented SAM-UNETR~\cite{alzate2023sam}, a FM adaptation for PCa detection in MRI, and MedSAM-UNETR, which substitutes SAM-UNETR's backbone with MedSAM. We additionally tested an adapter-tuned version of MedSAM, as well as Cinepro~\cite{harmanani2025cinepro}, a method designed for conventional ultrasound. \textit{Ablations:} We test the efficacy of the prompting strategy and by comparing \emph{no prompting} to different combinations of clinical metadata used as prompts. We test the efficacy of the multi-head class and mask decoder architecture by comparing it to single head versions.

\textbf{Prospective Validation}: Following model development and tuning with the retrospective data, we choose the configuration with the best cross-validation performance and freeze it. We then un-blind ourselves to the prospective data and pathology labels and evaluate the model, generating heatmaps and csPCa predictions. We assess the performance of these prospective predictions.

\textbf{Evaluation Metrics:} We assess core-level classification, measuring AUROC, sensitivity and specificity.\footnote{In retrospective evaluation, sensitivity is measured at multiple specificity levels; in prospective evaluation, a decision threshold is chosen and metrics are reported based on that threshold.} We assess performance for PCa (GG2+) vs. non-PCa \emph{and} csPCa (GG3+) vs. non-csPCa classification. For the former, we use the average heatmap pixel intensity in the needle trace region as a PCa score. For the latter, we use the \emph{model risk score}. 
We qualitatively evaluate our model by overlaying heatmaps on corresponding input US images. 
Following ~\cite{kinnaird2025microultrasonography}, we also assign each patient an overall PRI-MUS, PI-RADS and \emph{model risk score} by choosing the maximum score assigned to any of their biopsy images, and compare this score to the patient-level diagnosis.

\textbf{Implementation:} Training was performed on a single NVIDIA RTX-6000 (24 GB) with PyTorch 2.1. The Adam optimizer was used with a learning rate of $1\times10^{-5}$ and cosine annealing schedule. Each of five cross-validation folds are trained for 35 epochs (batch size 8, zero weight decay) and finish in $\approx 1.5$ h. Random image translation was used for data augmentation. Inference throughput is $27.6$ FPS on GPU and $2.63$ FPS on CPU. Full code and configurations are in our \href{https://github.com/pfrwilson/prostnfound}{public repository}.

\section{Results}

\subsection{Retrospective Evaluation}

\begin{table}[t]
\caption{Retrospective performance comparison across baseline and proposed methods. Averages and standard deviation across five folds are reported.}
\label{tab:retro_metrics}
\setlength{\tabcolsep}{4pt} 
\begin{tabular*}{\textwidth}{@{\extracolsep\fill}lccccc}
\toprule
\multirow{2}{*}{\textbf{Method}} & 
\multirow{2}{*}{AUROC$\uparrow$ (\%)} & 
\multicolumn{3}{c}{Sensitivity $\uparrow$ (\%)} \\
\cmidrule{3-5}
 & & 20\% Spe. & 40\% Spe. & 60\% Spe. \\

\midrule
USFM-UNETR~\cite{jiao2024usfm} & $61.2 \pm 3.0$ & $89.7 \pm 1.3$ & $74.9 \pm 2.2$ & $55.3 \pm 6.6$ \\
SAM-FT & $58.7 \pm 5.4$ & $88.2 \pm 4.1$ & $72.3 \pm 8.1$ & $52.6 \pm 7.4$ \\
SAM-UNETR~\cite{alzate2023sam} & $71.4 \pm 4.0$ & $ 92.8 \pm 2.3$ & $84.8 \pm 4.3$ & $71.3 \pm 4.4$ \\  
MedSAM-UNETR~\cite{alzate2023sam} & $69.7 \pm 3.4$ & $90.4 \pm 2.5$ & $81.1 \pm 3.6$ & $68.6 \pm 4.4$ \\
MedSAM-FT~\cite{ma2024segment} & $71.2 \pm 2.0$ & $95.2 \pm 1.7$ & $84.2 \pm 1.7$ & $70.2 \pm 2.6$ \\
Cinepro~\cite{harmanani2025cinepro} & $71.0 \pm 3.9$ & $93.4 \pm 3.8$ & $84.0 \pm 3.6$ & $70.3 \pm 3.8$  \\
\midrule
Patch-ResNet~\cite{wilson2023self} & $68.0 \pm 3.6$ & $93.3 \pm 1.3$ & $83.2 \pm 5.5$ & $66.9 \pm 5.5$ \\
MicroSegNet-FT~\cite{jiang2024microsegnet} & $73.9 \pm 0.9$ & $\mathbf{95.6} \pm 2.1$ & $85.4 \pm 1.8$ & $71.7 \pm 2.2$ \\

\midrule
ProstNFound~\cite{wilson2024prostnfound} & $76.6\pm 3.9$ & $95.5 \pm 2.4$ & $88.7 \pm 3.6$ & $77.7 \pm 6.0$ \\
\textbf{ProstNFound+} & $\mathbf{77.5} \pm 3.7$ & $95.3 \pm 2.5$ & $\mathbf{89.3} \pm 4.3$ & $\mathbf{79.3} \pm 5.5$\\
\bottomrule
\end{tabular*}
\end{table}

\textbf{Baseline Comparison:} Results are shown in Table~\ref{tab:retro_metrics}. For FM adaptations (Row Group 1), performance varies significantly, with MedSAM-based models (MedSAM-UNETR, MedSAM-FT and Cinepro) consistently performing well.  
Methods specifically developed for $\mu$US (Row Group 2) have mixed performance: while PatchResNet fails to outperform the FM adaptations, MicroSegNet-FT results in an AUROC of 73.9 (the second best), suggesting that its pretraining in $\mu$US-based prostate segmentation provides good transfer learning for PCa detection. ProstNFound+, which is based on MedSAM but also includes auxiliary prompting and a multi-head setup, achieves the best overall performance across most metrics, with its AUROC of $77.5\%$ being the highest by a substantial margin, and leading in sensitivity at most thresholds.

\begin{figure}
    \centering
    \includegraphics[width=.95\linewidth]{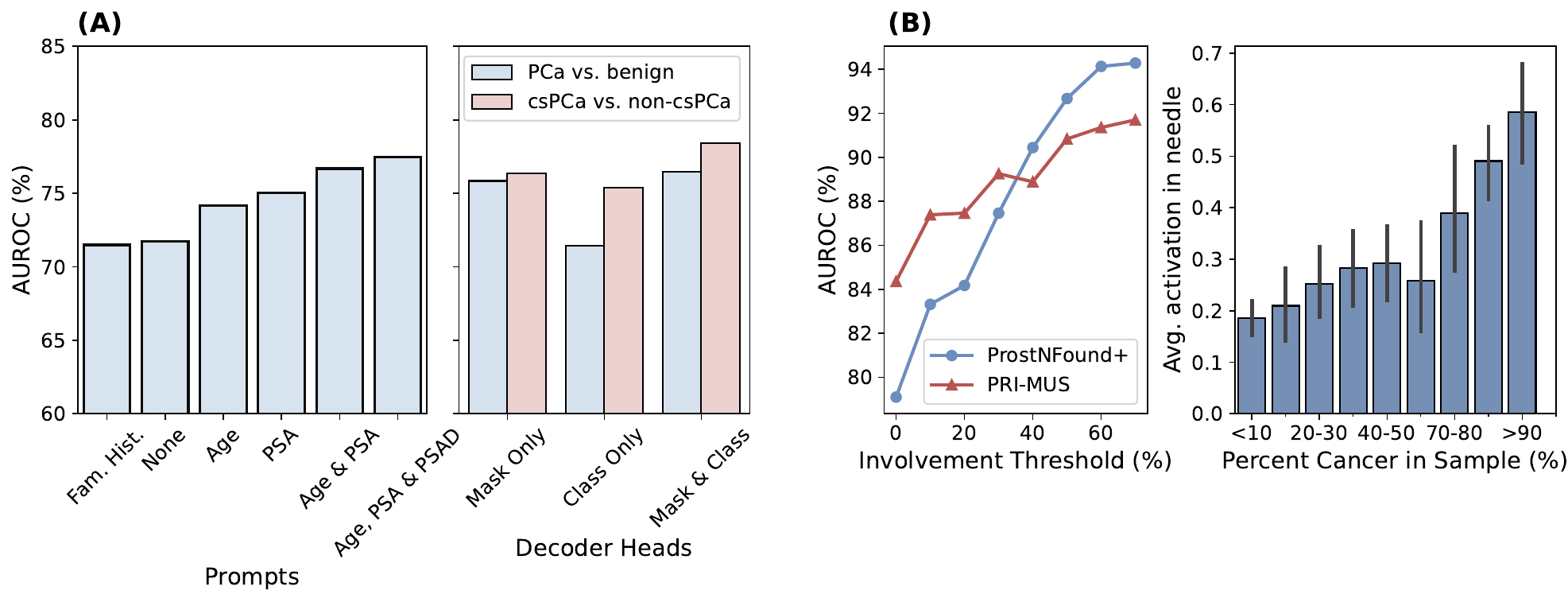}
    \caption{\textbf{(A)} Ablation study results. \textbf{(B)} csPCa detection performance (left) and average heatmap activation (right) by true involvement of cancer in samples.}
    \label{fig:combined_ablation_inv}
\end{figure}

\textbf{Ablation Studies:} 
Figure~\ref{fig:combined_ablation_inv}(A) summarizes results. Prompting with clinical features age, PSA, and PSAD improve AUROC, especially when combined (+$6\%$ over baseline). Since PSAD is not available for the prospective data, only age and PSA are used in the final model. 
A combination of mask and class decoders improved both csPCa–non-csPCa and PCa–benign classification compared to using one component only, indicating that they complement each other for cancer localization and aggressive disease detection. These ablations support the design choices of ProstNFound+.

\subsection{Prospective Evaluation}

\textbf{Biopsy-level results:} 
Table~\ref{tab:optimum_comparison} shows classification performance for ProstNFound+ compared to the visual protocols PRI-MUS and PI-RADs. For the full 1031-samples dataset, ProstNFound+ has an AUROC of $79.1\%$ compared to $84.4\%$ for PRI-MUS for csPCa detection and $74.2\%$ compared to $77.5\%$ for PCa detection. ProstNFound+ has similar specificity but lower sensitivity (-$10\%$) than PRI-MUS for csPCa detection. On the subset of samples with PI-RADS scores, ProstNFound+ had only $1\%$ lower AUROC than PI-RADS and $3.5\%$ lower AUROC than PRI-MUS for detecting csPCa. 

Figure~\ref{fig:combined_ablation_inv}(B) depicts the relationship between the cancer-in-core involvement and model performance. Both PRI-MUS and ProstNFound+ have improved performance on higher-involvement cores, with ProstNFound+ overtaking the performance of PRI-MUS for involvement at least $40\%$. (left plot). We hypothesize that higher involvement is correlated with larger tumors which are easier to detect. Furthermore, we find that higher-involvement cores are associated with larger numbers of pixels activated in the needle region on the corresponding heatmap (right plot), suggesting that ProstNFound+ can localize cancers within images.

\begin{table}[t]
\caption{Biopsy-level classification on the prospective dataset, comparing ProstNFound+ to expert-level visual scores of  PRI-MUS ($\mu$US) and PI-RADS (mpMRI).}
\centering
\begin{tabular*}{\textwidth}{@{\extracolsep\fill}lccc|ccc@{}}
\toprule
\multirow{2}{*}{\bf Method} & \multicolumn{3}{c|}{\bf csPCa (GG3+) vs. non-csPCa} & \multicolumn{3}{c}{\bf PCa (GG2+) vs. non-PCa} \\
 & AUROC (\%) & Sens. (\%) & Spec. (\%) & AUROC (\%) & Sens. (\%) & Spec. (\%) \\
\midrule
\multicolumn{7}{c}{{All biopsy samples ($n=1031$)}} \\
PRI-MUS         & 84.4 & 85.1 & 71.4 & 77.5 & 71.8 & 75.4 \\
ProstNFound+    & 79.1 & 75.2 & 70.3 & 74.2 & 73.7 & 63.5 \\
\midrule
\multicolumn{7}{c}{{Biopsy samples with PI-RADS scores ($n=308$)}} \\
PRI-MUS         & 91.6 & 100  & 73.5 & 87.4 & 88.3 & 79.9 \\
PI-RADS         & 88.1 & 92.8 & 74.8 & 84.0 & 81.4 & 80.7 \\
ProstNFound+    & 87.1 & 85.7 & 73.1 & 79.1 & 76.7 & 67.8 \\
\bottomrule
\end{tabular*}
\label{tab:optimum_comparison}
\end{table}

\begin{figure*}[htbp]
    \centering
    \includegraphics[width=0.88\textwidth]{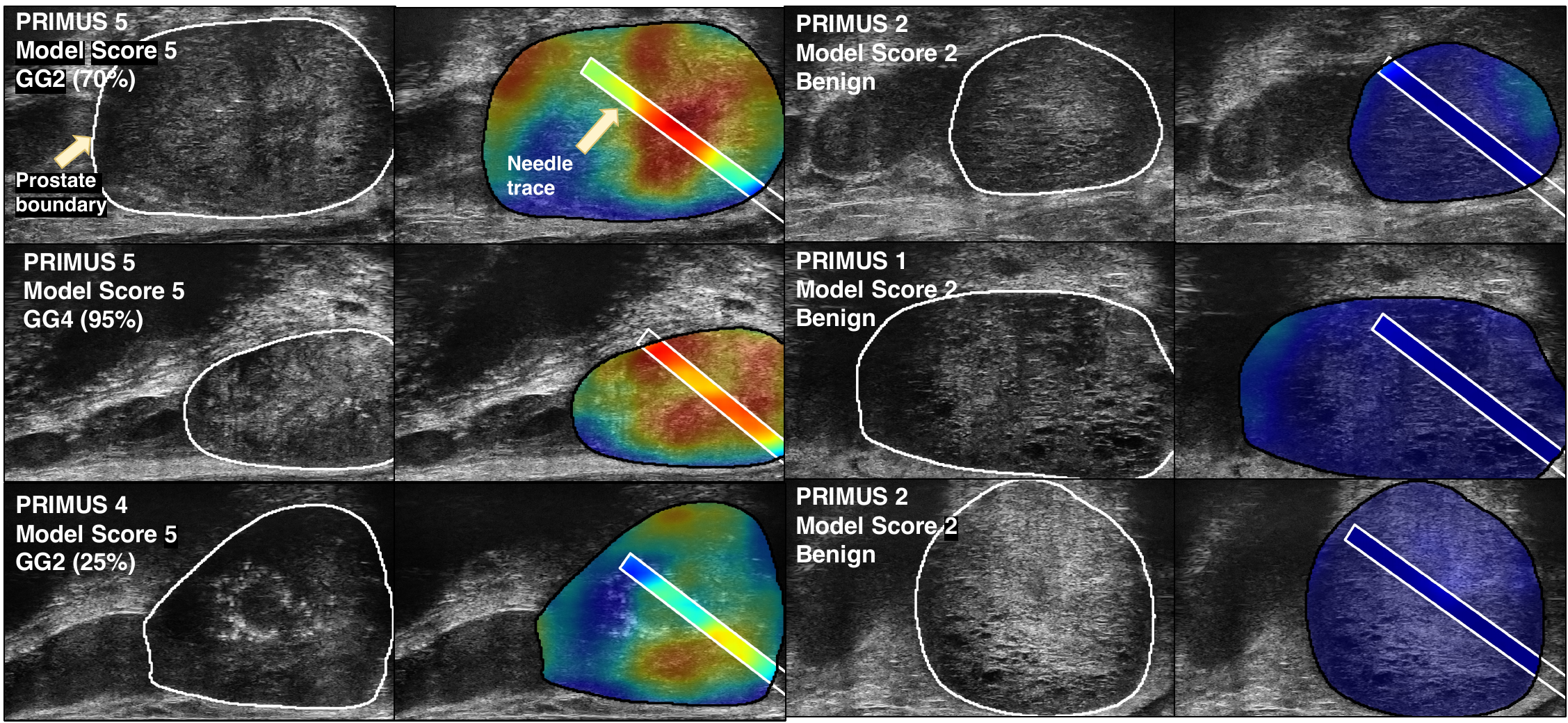}
    \caption{Example heatmaps generated by ProstNFound+. Higher PRI-MUS and \emph{model risk scores} indicate higher suspicion of cancer, and red activations indicate suspicious lesions. Suspicion of cancer typically coincides with true cancer confirmed by biopsy.}\label{fig:prostate_hmaps}
\end{figure*}

\textbf{Heatmap analysis:} Figure~\ref{fig:prostate_hmaps} shows examples of ProstNFound+'s predictions overlaid as heatmaps. In most benign cases (right), the model has low activations and low \textit{model risk score} indicating low suspicion. For csPCa cases (left), the model assigns high risk scores and highlights suspicious lesions in red. In deployment, targeted biopsy could be performed by acquiring biopsy samples from these lesions.

\begin{figure}[t] 
    \centering
    \includegraphics[width=\textwidth]{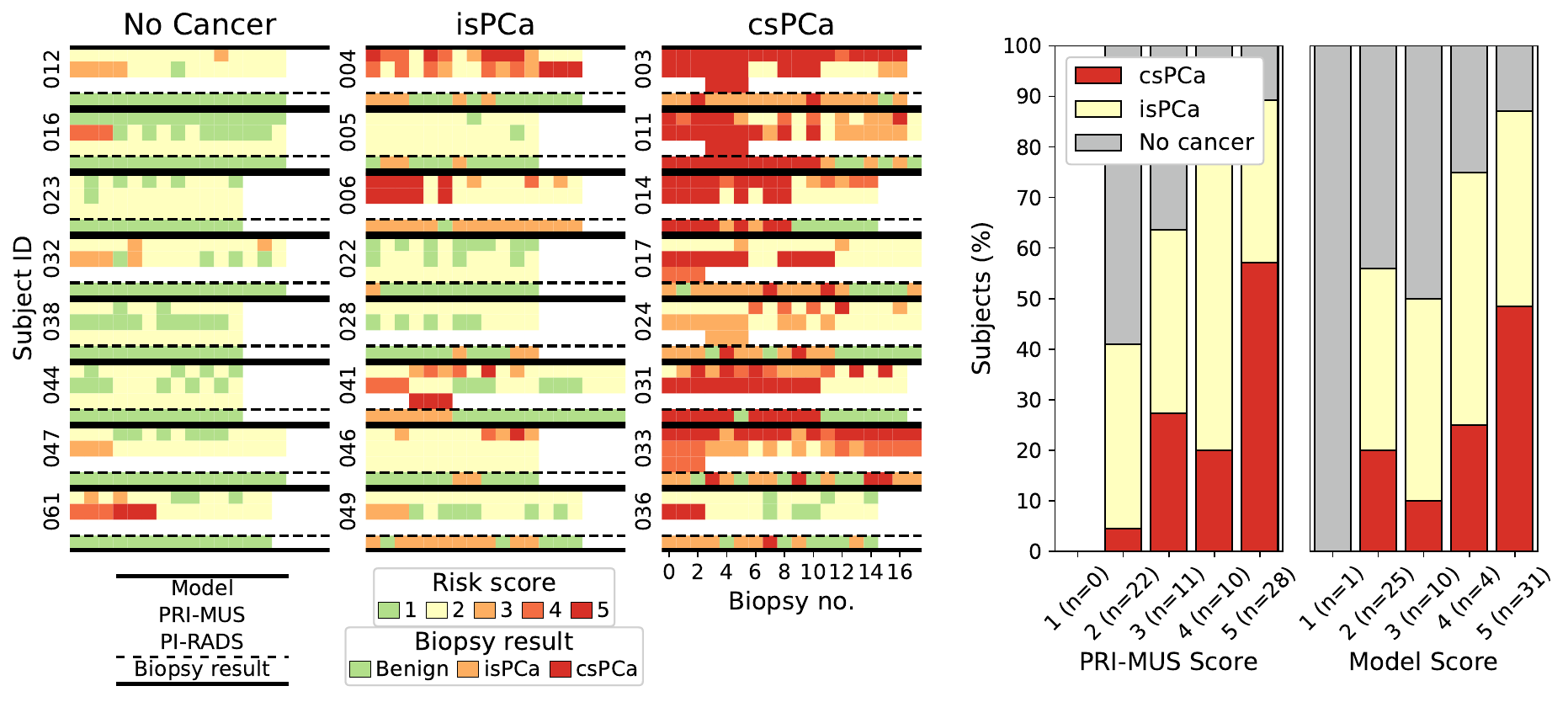}
    \caption{Left: Risk scores and results across biopsies for subjects in the prospective test set. Row groups distributed vertically represent subjects, and horizontal position within row group represents biopsy number. Subjects are grouped into columns by subject-level diagnosis. Right: breakdown of subject-level diagnoses (\emph{no cancer}, \emph{isPCa} \emph{csPCa}) by highest PRI-MUS score and highest \emph{model risk score} for that patient.}
    \label{fig:checkerboard}
\end{figure}

\textbf{Patient-level analysis:} A breakdown of the ProstNFound+, PRI-MUS, and PI-RADS risk scores together with biopsy results, by biopsy sample and by subject are shown in Figure~\ref{fig:checkerboard} (left side). Results are shown for 8 patients with No cancer, isPCa and csPCa diagnoses.
In general, scores align between scoring systems and match biopsy results. For patients with no cancer, PRI-MUS, PI-RADS and \emph{model risk score} are low ($\leq2$) with few exceptions. For patients with csPCa, scores are much higher in general: importantly, in most csPCa cases shown, \emph{at least one} csPCa core is assigned a model score of 5 (the highest score), suggesting that biopsies guided by \emph{model risk score} would likely detect csPCa at the patient level even if it does not detect it in every biopsy sample.       
Figure \ref{fig:checkerboard} (right side) shows that subjects with csPCa diagnosis (red boxes) or isPCa core (yellow boxes) were more likely to be assigned higher \emph{model risk scores} and PRI-MUS scores than patients with no cancer (grey boxes). A subject-level \emph{model risk score} or PRI-MUS score of 5 was highly specific for cancer, with only $10\%$ of subjects with that score not having cancer.

\section{Discussion}

In this study, the ProstNFound+ model successfully generalized from the retrospective dataset to a prospective dataset acquired at a different center 5 years later. No substantial decrease in any performance metric was observed between cross-validation and prospective test data. Due to shifts in operator technique, imaging hardware, and patient populations, such generalization was not guaranteed; this success is encouraging and increases confidence that deep learning models can perform well in clinical deployment. Trials directly assessing biopsy guidance accuracy and downstream clinical outcomes are needed to confirm efficacy.

Compared to the PRI-MUS scores, our model has a relatively small performance gap, with approximately $5\%$ lower AUROC. In some cases, it matches or exceeds the performance of PRI-MUS (e.g., when considering biopsies with at least $40\%$ cancer, corresponding to larger tumors). 
Given the extensive training required to achieve PRI-MUS expertise, the model has great potential value despite this performance gap, especially in cases where (i) PRI-MUS trained urologists are unavailable or have limited confidence and experience; or (ii) as a complementary user-agnostic tool to use in combination with more subjective visual methods. 

With continued improvements in deep learning methodology, models may soon close the performance gap with PRI-MUS and PI-RADS systems. Future efforts for improvement could address the following limitations: (i) The model was trained on core-level biopsy labels, which cannot provide accurate pixel-level labeling, contributing to its inability to localize small lesions. Training on pixel-level annotated images could improve this. Such annotations could be obtained by identifying biopsy-confirmed lesions and clearly delineating the tumour boundary, or by using coregistered whole mount pathology~\cite{pensa2024deep}. (ii) Our models only analyzed 2D images in isolation. PRI-MUS analysis depends on scanning the whole prostate gland to identify lesions, suggesting a benefit of developing 3D networks for PCa detection.

\section{Conclusion}

Our prospective validation study demonstrates that medical FMs like ProstNFound+ are approaching the diagnostic performance of established visual scoring systems and have the potential to generalize well in real-world clinical settings. Importantly, AI offers scalable, operator-agnostic analysis. Given the high inter-observer variability and learning curve of PRI-MUS, automated tools may offer a more consistent and accessible solution for $\mu$US-guided prostate cancer diagnosis, particularly in low-resource settings where MRI or expert interpretation is limited.

\section*{Declarations}

This research was supported by Natural Sciences and Engineering Research Council of Canada (NSERC), Canadian Institutes of Health Research (CIHR), the Vector Institute 
and a Canada CIFAR AI Chair to P Mousavi. B
Wodlinger is Vice President of Clinical and Engineering at Exact Imaging. There are 
no other conflicts of interest. Data used were from clinical trials approved by ethics boards and where informed consent was given for the use of data.



\bibliography{sn-bibliography}

\end{document}